\documentclass[3p,times,procedia]{elsarticle}
\usepackage{nupha_ecrc}

\volume{00}

\firstpage{1}

\journalname{Nuclear Physics A}

\runauth{K.~Nagai et al.}

\jid{nupha}

\jnltitlelogo{Nuclear Physics A}

\usepackage{amssymb}

\usepackage[figuresright]{rotating}

\def\del{\mathrm{d}}

\begin{document}

\begin{frontmatter}



\dochead{}

\title{Causal hydrodynamic fluctuation in Bjorken expansion}


\author[label1]{K.~Nagai}
\author[label1,label2,label3]{R.~Kurita}
\author[label1,label2,label3]{K.~Murase}
\author[label1]{T.~Hirano}
\address[label1]{Department of Physics, Sophia University, Tokyo 102-8554, Japan}
\address[label2]{Department of Physics, the University of Tokyo 113-0033, Japan}
\address[label3]{Theoretical Research Division, Nishina Center, Riken, Wako 351-0198, Japan}

\begin{abstract}
We investigate effects of causal hydrodynamic fluctuations
in the longitudinally expanding quark gluon plasma  
on final entropy distributions in high-energy nuclear collisions.
\end{abstract}


\end{frontmatter}


\section{Introduction}
In high-energy nuclear collisions, the space-time evolution of the quark-gluon plasma (QGP)
is well described by relativistic hydrodynamics.
So far, a vast body of analyses have been done to extract the bulk and transport
properties of the QGP by using event-by-event viscous fluid-dynamical simulations.
In these simulations, initial fluctuation of the transverse profile has been mainly focused on
to describe the higher order harmonics of azimuthal momentum distributions of
observed hadrons.
Moreover, whether hydrodynamic description of the QGP can be applied
in small colliding systems such as proton-proton and/or proton (deuteron)-nucleus collisions
is one of the open problems for these years.

In this study, instead of the initial fluctuations, 
we investigate hydrodynamic fluctuations 
\cite{Calzetta:1997aj, Kapusta:2011gt,Kumar:2013twa,Murase:2013tma}
to consistently describe the space-time evolution of the QGP 
on an event-by-event basis.
We focus on 
the boost-invariant Bjorken expansion \cite{Bjorken:1982qr}
within the framework of causal hydrodynamic fluctuations
to study the effect of them on final entropy distributions.

Throughout this contribution, we employ natural unit $c = \hbar = k_{\mathrm{B}} = 1$.

\section{Models}
We assume the system created in high-energy nuclear collisions
expands in a boost-invariant way in the beam direction.
The four velocity in this case can be parametrised as \cite{Bjorken:1982qr}
\begin{equation}
   u^{\mu} =\left(\cosh \eta_{\mathrm{s}}, 0, 0, \sinh\eta_{\mathrm{s}} \right),
\end{equation}
where  $\eta_{\mathrm{s}}=\frac{1}{2} \ln[(t+z)/(t-z)]$ is space-time rapidity.
We neglect baryon density at relativistic energies.
Then hydrodynamic equations reduce to
\begin{equation}
\label{eq:e}
\frac{\del e}{\del \tau}=-\frac{e+P}{\tau}\left(1-\frac{\pi}{sT}+\frac{\Pi}{sT}\right),
\end{equation}
where $\tau$ is proper time, 
$e$ is energy density, $P$ is hydrostatic pressure, 
$s$ is entropy density and $T$ is temperature.
 $\pi = \pi^{00}-\pi^{33}$ is an quantity calculated from
shear stress tensor $\pi^{\mu \nu}$ in Bjorken expansion and $\Pi$  is bulk pressure.
Time evolution of these dissipative quantities
are described by the constitutive equations.
Within causal hydrodynamic fluctuations,
these equations in the differential form can be written as
\begin{eqnarray}
\label{eq:pi}
\tau_{\pi}\frac{\del \pi}{\del \tau}+\pi & = & \frac{4\eta}{3\tau}+\xi_{\pi},\\
\label{eq:Pi}
\tau_{\Pi}\frac{\del \Pi}{\del \tau}+\Pi & = & -\frac{\zeta}{\tau}+\xi_{\Pi}.
\end{eqnarray}
Here $\eta$ and $\zeta$ are viscous coefficients and
$\tau_{\pi}$ and $\tau_{\Pi}$ are the corresponding relaxation times \cite{Israel} for shear viscosity
and bulk viscosity, respectively.
In the relativistic dissipative hydrodynamics,
there is an issue on causality within the first order theory.
So the relaxation terms in the left hand side of Eqs.~(\ref{eq:pi}) and (\ref{eq:Pi})
are crucial to obey the causality.
$\xi_{\pi}$ and $\xi_{\Pi}$ are Gaussian white noises for shear stress
and bulk pressure, respectively.
Statistical properties of these noise terms are determined from
transport coefficients through
the fluctuation-dissipation relations.
In the case of Bjorken expansion, the fluctuation-dissipation relations 
in the discretised version of the stochastic differential equations, 
(\ref{eq:pi}) and (\ref{eq:Pi}), are
\begin{eqnarray}
\langle\xi_{\pi}(\tau_{i})\xi_{\pi}(\tau_{j})\rangle&=&\frac{8T\eta\delta_{ij}}{3\Delta\tau\Delta{V}},\\
\langle\xi_{\Pi}(\tau_{i})\xi_{\Pi}(\tau_{j})\rangle&=&\frac{2T\zeta\delta_{ij}}{\Delta\tau\Delta{V}},\\
\langle\xi_{\pi}\rangle\, =\, \langle\xi_{\Pi}\rangle&=&0, \quad \langle\xi_{\pi}\xi_{\Pi}\rangle\, =\ 0.
\end{eqnarray}
Here $\Delta \tau$ is time step, $\Delta V =\tau\Delta\eta_{s}\Delta{x}\Delta{y}$ is
volume of a fluid element and angle bracket means ensemble average. 
It is noted that variances of noises are inversely proportional to the volume of a fluid element and, therefore, that the effect of hydrodynamic fluctuations
could be important in small colliding systems.

As for the equation of state, we first parametrise entropy density as a function of temperature
\cite{Asakawa:1995zu}
\begin{eqnarray}
s(T) & = &  d_{\mathrm{H}}\frac{4 \pi^2}{90} T^3 [1-f(T) ] +d_{\mathrm{Q}} T^3 f(T), \\
f(T) & = & \frac{1}{2} \left[1+\tanh \left(\frac{T-T_{\mathrm{c}}}{\Gamma} \right) \right].
\end{eqnarray}
The other thermodynamic quantities are calculated from
\begin{equation}
P(T) = \int_{0}^{T} s(T') dT', \quad e(T) = Ts(T) - P(T).
\end{equation}
Parameters $d_{\mathrm{H}} = 3$ and $d_{\mathrm{Q}} = 37$ 
are degrees of freedom in the hadronic matter and the quark gluon plasma, respectively.
Transition temperature is $T_{\mathrm{c}} = 0.17$ GeV and a width of transition is 
$\Gamma = 0.02 T_{\mathrm{c}}$. 
The equation of state obtained in this way exhibits the cross-over behaviour.

We employ models for transport coefficients as $\eta/s = 1/4 \pi$ \cite{Kovtun:2004de}
and $\zeta/s = 15 \left(1/3 - c_{\mathrm{s}}^2\right)$ \cite{Weinberg:1971mx}. 
Relaxation times are $\tau_\pi = \tau_\Pi = 3\eta/2p$.
It should be emphasised that the purpose of the present study is
to demonstrate the importance of hydrodynamic fluctuations
for shear and bulk viscosities and the choice of these 
models is not intended for any specific matter nor situation.

\section{Results}

In the following, we start from a fixed set of initial conditions.
Initial temperature is $T_{0} = 0.22$ GeV at initial time $\tau_0 = 1$ fm.
At the initial time dissipative quantities are assuemed to be vanishing.
As a typical size of one fluid element, we choose $\Delta x =\Delta y = 1$ fm
and $\Delta \eta_{\mathrm{s}} = 1$.	
Time step in the simulations is $\Delta \tau = $ 0.1 fm. 

\begin{figure}[htb]
  \includegraphics[width=0.35\textwidth,angle=-90]{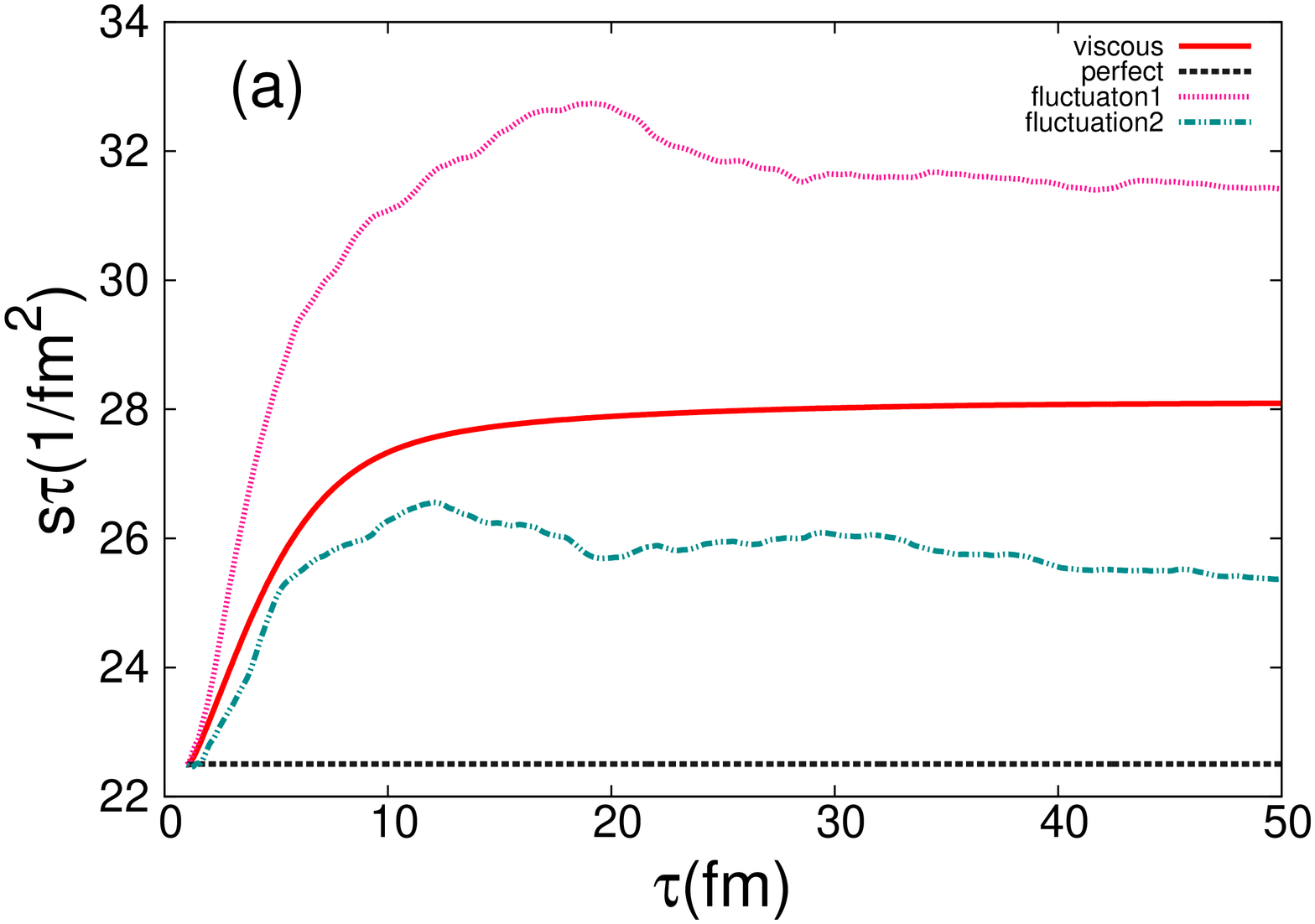}
  \includegraphics[width=0.35\textwidth,angle=-90]{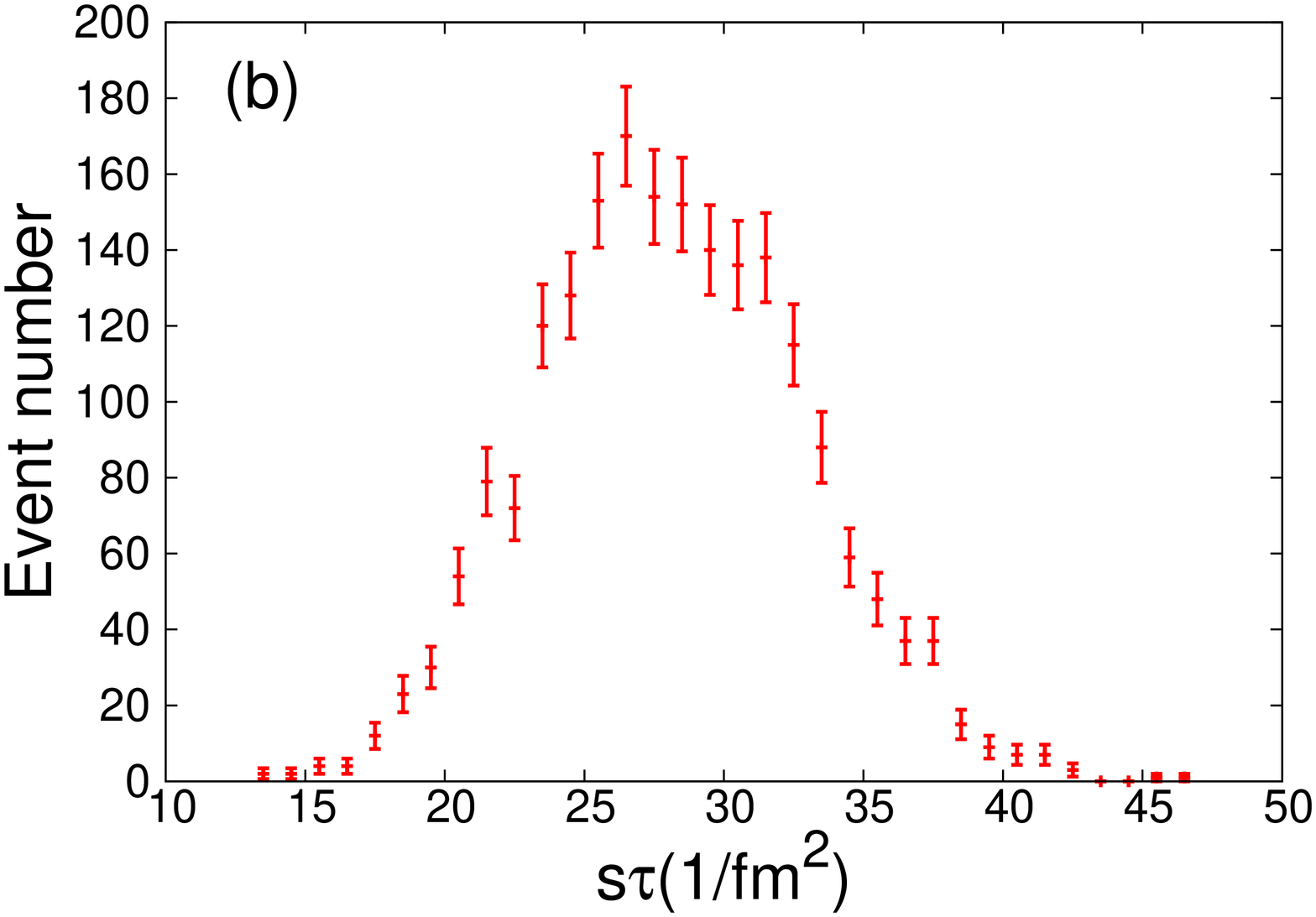}
  \caption{(a) Time evolution of entropy
for a perfect fluid (dashed), a viscous fluid (solid) and fluctuating fluids (dotted and dash-dotted).
(b) Histogram of the entropy distribution at proper
time $\tau= 50$ fm}
\label{fig:one}
\end{figure}

Time evolution of the product of the entropy density, $s$, and the proper time, $\tau$, 
is shown in Fig.~\ref{fig:one} (a).
When one neglects dissipative quantities
in Eq.~(\ref{eq:e}), the quantity $s\tau$
is the entropy per unit space-time rapidity and per unit transverse area
and a constant of motion for a perfect fluid in one-dimensionally expanding system
 \cite{Bjorken:1982qr}.
On the other hand,
$s\tau$ increases monotonically for a viscous fluid (neglecting noise terms
in Eqs.~(\ref{eq:pi}) and (\ref{eq:Pi})) due to production of entropy.
Since thermodynamic force is the inverse of the
proper time in the Bjorken expansion,
the production of entropy is dominated in the early stage ($<  10$ fm)
and the quantity $s\tau$ saturates in the late stage ($\sim 50$ fm).
We also show two results of causal hydrodynamic fluctuations
in Fig.~\ref{fig:one} (a).
Due to the noise terms in Eqs.~(\ref{eq:pi}) and (\ref{eq:Pi}),
the quantity $s \tau$ is fluctuating during the evolution
and final entropy is not the same in these two results
even though the initial conditions are common.
Interestingly, $s \tau$ can decrease temporarily during the time evolution.
This behaviour is explained by the entropy production rate $\sigma$
in the case of causal hydrodynamic fluctuation.
From Eqs.~(\ref{eq:e}) - (\ref{eq:Pi}), 
one derives entropy production rate in a fluid element,
\begin{eqnarray}
\sigma & = & \frac{\del \Delta S}{\del \tau}=\left(\frac{\pi}{T}-\frac{\Pi}{T}\right)
\Delta\eta_{\mathrm{s}}\Delta{x}\Delta{y}, \\
\Delta S & = & s \tau \Delta\eta_{\mathrm{s}}\Delta{x}\Delta{y}.
\end{eqnarray}
In the case of causal hydrodynamic fluctuations, shear stress and bulk pressure 
can take both positive and negative values due to noise terms.
As a result, entropy production rate sometimes takes negative values.
This is dynamical manifestation of the fluctuation-dissipation relation:
Fluctuations make the system deviate from the maximum entropy state
by reducing entropy, but dissipations move it back to the state
by producing entropy. 

We analysed two thousand events in causal hydrodynamic fluctuations.
Figure \ref{fig:one} (b) shows a histogram of the quantity $s \tau$ 
at $\tau = 50$ fm.
The shape of this entropy distribution looks like a Gaussian.
The mean value (28.06(15)/fm$^2$) is almost the same as the one for the viscous fluid (28.05 /fm$^2$).
When the effect of hydrodynamic fluctuations is sufficiently large,
one sees the final entropy being smaller than the initial entropy 
($s_0 \tau_0 = 22.49$ /fm$^2$)
in a small fraction of the total events in Fig.~\ref{fig:one} (b).
We emphasise here that this is not in contradiction with the second law of
thermodynamics. In fact, 
 one is able to quantify
the fraction of net negative entropy production 
from the fluctuation theorem \cite{Hirano:2014dpa}.

\begin{figure}[htb]
\centering
 \includegraphics[width=5.0cm,angle=-90]{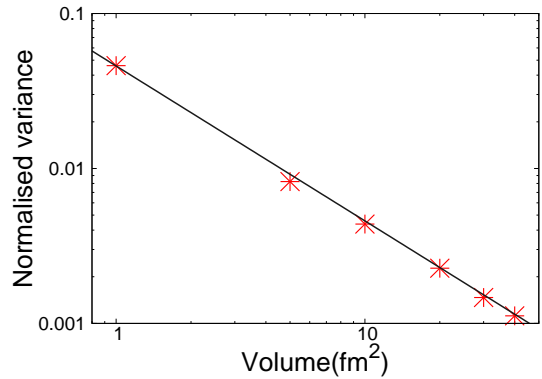}
 \caption{Volume dependence of variance of final entropy normalised by initial entropy.
The solid line is the inverse of the volume.}
\label{fig:two} 
\end{figure}

To see the volume dependence of hydrodynamic fluctuations,
we calculate variance of the final entropy distribution
normalised by initial entropy,
\begin{equation}
\frac{\langle{(s\tau-\overline{s\tau})}^2\rangle}{{(s_{0}\tau_{0})}^2}.
\end{equation}
Here $\overline{s \tau}$ is the average of the final entropy. 
Figure \ref{fig:two} shows a normalised variance of the final entropy distribution
as a function of volume ($=\Delta \eta_{\mathrm{s}}\Delta{x\Delta{y}}$). 
 As expected from Eqs.~(\ref{eq:pi}) and (\ref{eq:Pi}),
the normalised variance is inversely proportional to the volume.
Within the present setting of initial conditions,
an width of the final entropy distribution is a few percent of the initial entropy
in a moderate size of the volume ($\sim 1$ fm$^2$).
The smaller the size of a fluid element is, the more significant  the effect
of hydrodynamic fluctuations is. 
One would say 
that the effect of  hydrodynamic fluctuation
is important in the small colliding systems
such as proton-proton, proton (deuteron)-nucleus and, perhaps, 
peripheral nucleus-nucleus collisions.

\section{Summary}
We first formulated 
relativistic causal hydrodynamics with 
the effects of thermal fluctuations during the evolution 
(hydrodynamic fluctuations)
in the one-dimensionally expanding QGP.
We calculated entropy production within this framework and compare them with
that from the conventional viscous fluid-dynamics.
Final entropy fluctuates around the mean value due to hydrodynamic
fluctuations even when the system 
starts from the same initial condition in a macroscopic sense.
The entropy can temporally decreases during the evolution,
which can be understood from the fluctuation theorem in non-equilibrium statistical physics.
Hydrodynamic fluctuations should be implemented
 in the hydrodynamic description of the QGP on an event-by-event basis since,
through the fluctuation-dissipation relation,
the hydrodynamic fluctuations are always accompanied by the corresponding dissipations
in viscous fluid-dynamics.
Since multiplicity is almost proportional to the final entropy of the system in high-energy nuclear collisions, the effects of hydrodynamic fluctuations would be important in
multiplicity distributions in small colliding system.


 \appendix


This work was supported by JSPS KAKENHI Grant Numbers 
12J08554 (K.M.) and 25400269 (T.H.).



\bibliographystyle{elsarticle-num}
\bibliography{<your-bib-database>}



\end{document}